\documentclass[modern]{aastex61}

 \usepackage{natbib}                                                                                                               
 \usepackage{graphicx}

\newcommand{\etal}{{\it et al.}}
\newcommand{\ie}{{\it i.e.}~}
\newcommand{\eg}{{\it e.g.},~}

\newcommand{\Rmnum}[1]{{\uppercase\expandafter{\romannumeral #1}}}

\usepackage{epstopdf}
\usepackage{amsmath}

\newcommand{\Hunit}{ergs~cm$^{-2}$~DN$^{-1}$}
\newcommand{\etaunit}{cm~s$^{-1}$}
\newcommand{\SDO}{\textit{SDO}}

\newcommand{\Hinode}{\textit{Hinode}}
\newcommand{\GOES}{\textit{GOES}}
\hypersetup{
    colorlinks=true,       
     linkcolor=blue,          
     citecolor=blue,        
     filecolor=black,      
      urlcolor=blue           
}

\shorttitle{Two-phase Heating in Flaring Loops}
\shortauthors{Zhu \etal}

\begin{document}
\ifx \doiurl    \undefined \def \doiurl#1{\href{http://dx.doi.org/#1}{\textsf{#1}}}\fi
\ifx \adsurl    \undefined \def \adsurl#1{\href{http://adsabs.harvard.edu/abs/#1}{\textsf{#1}}}\fi
\ifx \arxivurl  \undefined \def \arxivurl#1{\href{http://arxiv.org/abs/#1}{\textsf{#1}}}\fi

\title{Two-phase Heating in Flaring Loops}

\correspondingauthor{Chunming Zhu}
\email{chunming.zhu@montana.edu}

\author{Chunming Zhu}
\affiliation{Physics Department, Montana State University, Bozeman, MT 59717-3840, USA}

\author{Jiong Qiu}
\affiliation{Physics Department, Montana State University, Bozeman, MT 59717-3840, USA}

\author{Dana W. Longcope}
\affiliation{Physics Department, Montana State University, Bozeman, MT 59717-3840, USA}

\begin{abstract}

We analyze and model a C5.7 two-ribbon solar flare observed by \SDO, \Hinode\ and \GOES\ on 2011 December 26. The flare is made of many loops formed and heated successively over one and half hours, and their foot-points are brightened in the UV 1600~\AA\ before enhanced soft X-ray and EUV missions are observed in flare loops. Assuming that anchored at each brightened UV pixel is a half flaring loop, we identify more than 6,700 half flaring loops, and infer the heating rate of each loop from the UV light curve at the foot-point. In each half loop, the heating rate consists of two phases, an intense impulsive heating followed by a low-rate heating persistent for more than 20 minutes. Using these heating rates, we simulate the evolution of their coronal temperatures and densities with the model of ``enthalpy-based thermal evolution of loops'' (EBTEL).  In the model, suppression of thermal conduction is also considered. This model successfully reproduces total soft X-ray and EUV light curves observed in fifteen pass-bands by four instruments \GOES, AIA, XRT, and EVE. In this flare, a total energy of $4.9{\times}10^{30}$ ergs is required to heat the corona, around 40\% of this energy is in the slow-heating phase. About two fifth of the total energy used to heat the corona is radiated by the coronal plasmas, and the other three fifth transported to the lower atmosphere by thermal conduction.

\end{abstract}


\section{Introduction}
Solar flares, observed as increased radiation across a broad band of electromagnetic spectrum, are generally accepted to be associated with a sudden release of free magnetic energy through the process of magnetic reconnection. During flares, the heated and accelerated particles travel along the newly formed coronal loops down toward the chromosphere, and deposit their energy at the loop footpoints, which usually form two evolving ribbons. The energy deposition there drives the chromospheric evaporation (\citealp{canfield1980chromosphere, fisher1984chromospheric}), which fills the coronal loops.  The heated coronal plasmas then cool down gradually due to thermal conduction and radiation (\citealp{culhane1970cooling, antiochos1978evaporative, cargill1995cooling}). 

The hydrodynamic evolution of the flaring plasmas has been investigated by many theoretical models. The properties and response of plasmas confined in coronal loops to some assumed heating mechanisms were studied by solving the one-dimensional (1D) hydrodynamic equations (\eg \citealp{mcclymont1983flare, nagai1984gas, longcope2010quantitative, Bradshaw2013}). However, the investigation of a wide range of parameters in various heating mechanisms (\citealp{Mandrini2000}) makes it very challenging for the computationally intensive 1D models. Thus the 0-dimensional (0D) models were developed to study the averaged values in each single loop/thread (\eg \citealp{Fisher1990,Kopp1993,Cargill1994}).  \citet{klimchuk2008highly} proposed an improved 0D model called ``enthalpy-based thermal evolution of loops'' (EBTEL) which gives an efficient way to calculate the average temperature and density in coronal loops/threads.

The response of the plasmas inside a coronal loop is governed by the energy input, or the heating rate. However, the physical mechanism of heating, and the amount of heating energy in flare loops, still remain largely unknown. \citet{qiu2012heating} proposed an intuitive empirical method to infer the heating rates in flare loops that are continuously formed throughout the flare, utilizing spatially resolved UV emission in the lower atmosphere. They assume that anchored at each newly brightened UV pixel is a flare (half) loop, and the impulsive rise of the UV light curve at the pixel is scaled with the heating rate in the loop. This is the so-called UV Footpoint Calorimeter (UFC) method.
With this method, hundreds to thousands of flare loops are identified in a flare even into the decay phase of the flare, when continuous energy release (and formation of new loops) still occur (\eg \citealp{cargill1983heating,czaykowska1999evidence,czaykowska2001chromospheric,Reeves2002}). With the inferred heating rates, \citet{qiu2012heating} and \citet{liu2013determining} compute the evolution of flare loops and synthesize SXR and EUV emissions therein, which compare favorably with observed emissions during the rise of the flare. Subsequently, \citet{qiu2016long} studied a flare that exhibits a long-duration emission at 10~MK and slow cooling to lower temperatures. They found that superposition of many intense impulsive heating events, even into the decay phase of the flare, cannot reproduce the observed signatures at different temperatures. To improve the model-observation agreement, they needed to use a two-phase heating profile for  {\em each flare loop or thread}: an intense impulsive heating, followed by a gradual slow heating. The two-phase profile may or may not coincide with a suppression of thermal conduction below its Spitzer value, in order to maintain the coronal plasma at high temperatures for a longer time (\eg \citealp{jiang2006evolution,battaglia2009observations,wang2015evidence}).

In this study, we analyze and model a two-ribbon flare with a modified UFC, and study the effects of two-phase heating as well as thermal conduction suppression (TCS) introduced in each flare loop. 
We find that the inclusion of both the persistent slow-heating and TCS in flare loops leads to the best agreement between model synthetic and observed SXR and EUV light curves in many pass-bands.
In Section 2, we give an overview of the C5.7 flare observed on 2011 December 26. In Section 3, we model the flare evolution with EBTEL and compare the synthetic X-ray and EUV light curves to the observations from \GOES, \SDO/AIA{\&}EVE,  and \Hinode/XRT. The energetics and physical properties of flare loops are analyzed in Section 4. Conclusions and discussions are given in Section 5.

\section{Overview of Observations}

This C5.7 flare was positioned northeast of an active region NOAA~11384, and near center of the solar disk.  We focus on the X-ray and Extreme-Ultraviolet (EUV) observations provided by three spacecraft including the {\it Solar Dynamics Observatory} (\SDO; \citealp{pesnell2012solar}),   \Hinode ~(\citealp{Kosugi2007}) and \GOES. \SDO ~has three observing instruments on board: the Atmospheric Imaging Assembly (AIA; \citealp{Lemen2012}) ~takes full-disk images of the Sun in 10 EUV/UV channels ($\log$T ranges 3.7--7.3) with roughly $0{\farcs}6$~pixel$^{-1}$ spatial resolution; the Helioseismic and Magnetic Imager (HMI; \citealp{Schou2012}) measures full-disk magnetograms with 1$''$ spatial resolution and 45-second cadence; and the Extreme ultraviolet Variability Experiment (EVE) provides irradiance with high spectral resolution.  The X-ray Telescope (XRT; \citealp{Golub2007}) on board \Hinode ~observes this flare during its early phase in multiple bandpasses with a scale of $\sim$1$''$ pixel$^{-1}$. \GOES ~has two X-ray sensors measuring the X-ray fluxes in the wavelength bands of 0.5--4~\AA\ (short channel) and 1--8~\AA\ (long channel).

\begin{figure*}[t]
  \begin{center}
        \includegraphics[viewport = 320 30 667 660, clip,angle=90, width=0.9\textwidth]{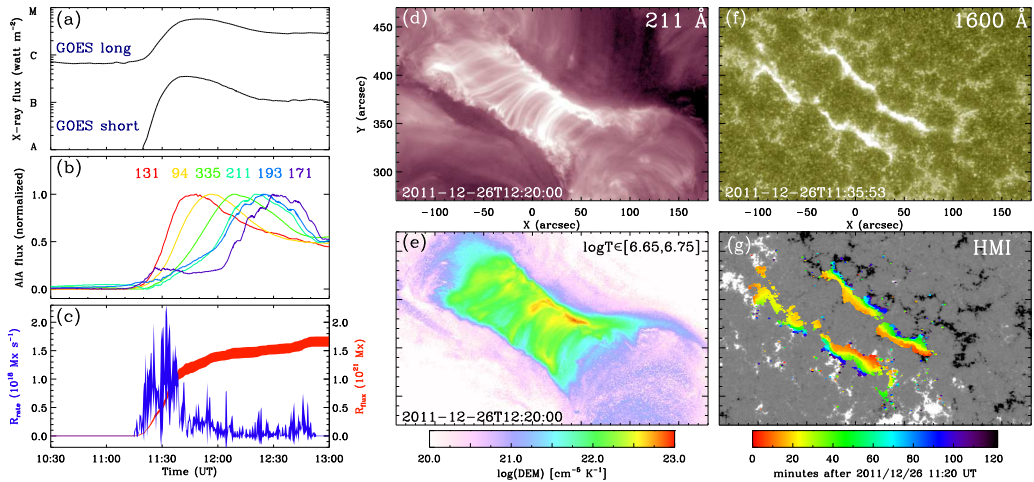}
        
    \caption{Overview of a C5.7 flare observed from \SDO\ on 26 December 2011.  (a\&b) lightcurves observed from \GOES\ X-ray and AIA EUV channels, respectively.  (c) Estimated reconnection rate (blue) and cumulative flux (red) when the flare ribbon expands, as shown in (g). (d) Flare loops observed in AIA 211~\AA. (e) Reconstructed differential emission measures (DEMs) with $\log$T between 6.65--6.75.  (f) Flaring ribbons observed in AIA 1600~\AA. (g) Expansion of the flare ribbons (1600~\AA, colored) which swept the magnetic fields (gray).}
    
    \label{fig:fig1}
  \end{center}
\end{figure*}

The \GOES\ soft X-ray in the long channel begins to increase at 11:23 UT and ends at 12:18 UT, with its peak appearing at 11:50 UT, as seen in Figure \ref{fig:fig1}(a). Figure \ref{fig:fig1}(b) gives the cooling process observed in the EUV channels from \SDO/AIA: the peaks of the lightcurves appear progressively from the hotter to cooler channels (\eg $\sim$10~MK in 131~\AA\ and 0.6~MK in 171~\AA).  Similar phenomena have been reported in previous studies (\eg \citealp{Ryan2013,Viall2013}).

The flaring loops, when its total peak brightness observed in AIA 211~\AA\ ($\sim$2 MK), are shown in Figure \ref{fig:fig1}(d).  The overall shape of these loops are usually well described as semi-circular (\eg \citealp{reale2014coronal}).  Six optically thin \SDO/AIA EUV channels (except 304~\AA) can be utilized to derive the emission measures at varying coronal temperatures.  Figure \ref{fig:fig1}(e) gives an example of the differential emission measures (DEMs) with $\log$T ranging from 6.65--6.75, calculated with the sparsity-based inversion method (\citealp{cheung2015thermal}). Similar DEM values appear along each loop in the flaring arcade, suggesting that evolution of flare loops, though formed and heated at different times, is rather similar.

Two elongated ribbons observed in AIA 1600~\AA\ are shown in Figure \ref{fig:fig1}(f). They are located beside the polarity inversion line, and spread outward sequentially, as seen in Figure \ref{fig:fig1}(g). The flare ribbons are composed of small kernels outlining the foot-points of flaring loops (\citealp{fletcher2004tracking}).  The distance between the two ribbons are increasing from 31 Mm at 11:27 UT to 42 Mm at 12:00 UT, indicating that loops anchored at newly brightened flare ribbons become longer, as magnetic reconnection forming these loops occurs at progressively higher altitudes (\eg \citealp{gallagher2003rhessi}). The reconnection rate (\eg \citealp{forbes1984numerical,qiu2004magnetic,kazachenko2017database}), estimated by the amount of magnetic flux swept by the flaring ribbons at a given time, is shown with the blue curve in Figure~\ref{fig:fig1}(c), with the cumulative flux in red.

\section{Modeling Plasma Evolution in Flaring Loops}

We use the UFC method, with some modifications, to infer heating rates and model evolution of flare loops. The AIA 1600 \AA\ images are processed using the standard routine aia\_prep and then differentially rotated to a time just before the flare at 11:00 UT. The brightening pixels in 1600 \AA\ are chosen with two criteria: [1] their values are larger than a threshold of $\sim$200 DN/s, which corresponds to 2.5 times of the median value of all pixels in the region of interest before the flaring, and [2] the brightening in each pixel lasts for at least 3 minutes. A few tests suggest that the outputs are not sensitive to the arbitrary values in both criteria. As a result, there are 6,700 brightening pixels in total identified in AIA 1600~\AA\ in this C5.7 flare. With each such a pixel we assume a half flaring loop of a constant cross-section ($0{\farcs}6{\times}0{\farcs}6$) is rooted in it. Then we investigate the evolution of the plasma parameters of each half loop with EBTEL. 

\subsection{EBTEL Setup}

The basics of setting up EBTEL can be found in \citet{qiu2012heating} and \citet{liu2013determining}. 
The model solves two equations, an energy equation and a mass conservation equation, to compute the time evolution of the mean temperature and density of a flare loop, assuming that the corona and transition region evolve in equilibrium, \ie uniform pressure. Energy input in the corona is required to run the model, and energy loss terms include radiations by the corona and transition region. During the heating phase, energy is transferred, such as by thermal conduction, from the corona to the transition region, which in turn transports mass (and enthalpy flux) back to the corona. 

To model the flare evolution, we first determine some loop properties from observations, the length of the loop and the heating rate in each loop. For this C-class flare without significant non-thermal emission above 20~keV, we do not consider heating by chromospheric evaporation driven by non-thermal particles that precipitate in the lower atmosphere; therefore, all corona heating is in-situ. In this paper, we use ad-hoc coronal heating rates inferred from UV light curves and do not explore the mechanism for the in-situ heating. Improved over the standard UFC method, we include TCS in the model, and also examine the effect of slow heating following the impulsive heating in each loop.

As the flare progresses, two ribbons separate indicating larger lengths of newly formed loops. We approximate the lateral expansion of the ribbons by a linear increase with time. The half-loop length of the flaring arcade also grows linearly from 24~Mm at 11:27~UT to 33~Mm at 12:00~UT,
described by $\mathrm{L=24+0.27(t-t_0)}$~Mm, where $t_0$ is the time of flare onset at 11:27~UT, and $t$ is the time of the peak UV brightening at the foot of the half loop expressed in minutes. We assume that the length of a particular half-loop does not change during its subsequent evolution. Before $t_0$ of 11:27 and after 12:00~UT, the lengths are fixed at 24 and 33~Mm, respectively.  

Under the flaring conditions, the thermal conduction can sometimes be suppressed (\eg \citealp{jiang2006evolution,battaglia2009observations,wang2015evidence}). In this study, we consider the TCS given by \citealp{rosner1985physical}, \ie when the ratio of the mean free path for thermal electrons $\mathrm{l_{mfp}}$ is larger than 0.015 of the temperature scale length $\mathrm{L_{th}}$ (here using the loop half length), a reduction factor of $\mathrm{0.11(l_{mfp}/L_{th})^{-0.36}}$ is applied to the classical thermal conduction (\citealp{spitzer1962physics}) until it is further saturated (\citealp{luciani1983nonlocal,Karpen1987}). Here we choose $\mathrm{l_{mfp} = 1.4{\times}10^7(T/10^6~ K)^2(n/10^9~cm^{-3})^{-1}~cm}$, where T and n are the average temperature and density in each loop, respectively. We adopt the same expressions of the classical and saturated thermal conductions as shown in equations 18--22 in \citet{klimchuk2008highly}.  
 
\begin{figure*}[t]
  \begin{center}
    \includegraphics[viewport =208 78 400 738, clip,angle=90,width=0.95\textwidth]	
    	{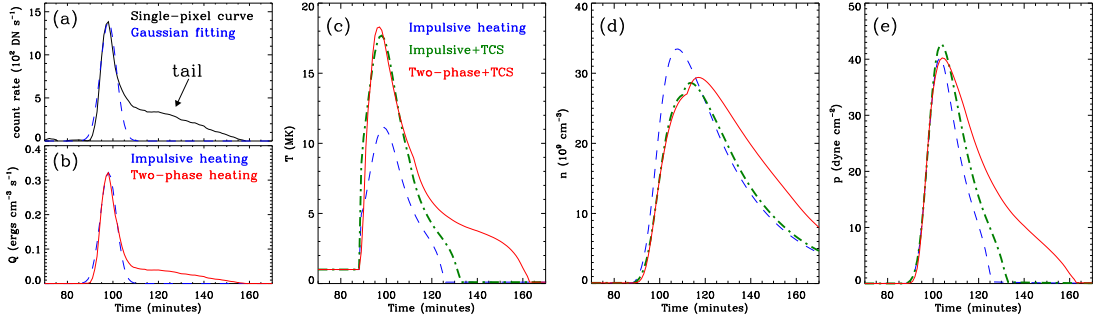}  			    	
    \caption{Heating and the resultant response of the plasma in one flaring loop. (a) Lightcurve of a single pixel in AIA 1600~\AA\ and its Gaussian fitting. The tail part is indicated. (b) Construction of the heating rates with two heating mechanisms: impulsive heating and two-phase heating. The impulsive heating function is based on the Gaussian fitting, while the two-phase heating has an additional slow tail that is proportional to the tail of the lightcurve, which is denoted in (a). (c--e): Evolution of temperature (T), density (n), and pressure (P) of this individual flaring loop given by EBTEL in three scenarios: impulsive heating (dashed), impulsive heating with thermal conduction suppression (TCS, dash-dotted), and two-phase heating with TCS (solid), respectively.}
    \label{fig:fig2}
  \end{center}
\end{figure*}

The heating rates are derived from the light-curves of the associated flaring pixels in AIA 1600 \AA.  The lightcurve of such a pixel is shown in Figure \ref{fig:fig2}(a). The standard UFC method fits the rise of the UV light curve with a half-Gaussian, and assumes that the impulsive heating flux is proportional to the full Gaussian, as indicated by the dashed line in the figure. The observed UV lightcurve typically decays much slower than its rise, with a gradually attenuated tail following the Gaussian fitting. The slow decay of the UV lightcurve may be partly due to continuous heating of the transition region by thermal conduction from the corona {\em without} more energy deposit into the corona; however, it is also likely that during this slow decay, additional heating also takes place in the corona. To understand the effect of slow heating during the decay, in this study, we model and compare flare loop evolution with two types of heating rates, impulsive heating and two-phase heating.  Following \citet{qiu2012heating}, the \textit{impulsive heating rate} $\mathrm{H_{imp}}$ is chosen to be proportional to the Gaussian fitting of the lightcurve with a scalar factor $\lambda_0$ in units of \Hunit,
which converts the UV count rates $\mathrm{I_{imp}}$ to the impulsive heating flux by $\mathrm{H_{imp} = \lambda_0 I_{imp}}$. 
The \textit{two-phase heating} contains an extra gradual heating $\mathrm{H_{grad}}$, which in this study is assumed to be proportional to the slow-tail of the UV light curve ($\mathrm{I_{tail}}$) by another scaling factor $\lambda_1$, having the same units as $\lambda_0$, \ie $\mathrm{H_{grad}=\lambda_1 I_{tail}}$. 
Such reconstructed heating functions are displayed in Figure \ref{fig:fig2}(b). The same values of  $\lambda_0$ and $\lambda_1$ are used for all loops. They are determined by comparing model synthetic SXR emission with that observed by \GOES.

The radiative loss from the transition region is also specified in the model as scaled with the mean pressure of the corona by a scaling constant $\eta$, which is the same for all loops. This parameter is chosen by comparing the model synthetic EUV emission at the low temperature (1--2~MK) with observations \citep{qiu2016long}.

Given the heating functions and the half-length of the flaring loop, its evolution can be modeled with EBTEL. We considered three scenarios: [\Rmnum{1}] impulsive heating, [\Rmnum{2}] impulsive heating with TCS, and [\Rmnum{3}] two-phase heating with TCS. Figures \ref{fig:fig2}(c--e) show the temperature, density, and pressure of one flare loop, with $\lambda_0 = 6.3{\times}10^5$ \Hunit, $\lambda_1 = 3.2{\times}10^5$ \Hunit, and $\mathrm{L = }$27.3 Mm,  modeled in these three cases. It is notable that: 
[1] TCS helps retain more energy in the corona and thus lead to a higher temperature; the suppressed conduction drives less chromospheric evaporation, therefore the peak density is lower; and the resulting effect leads to comparable pressures. 
[2] The slow tail in the two-phase heating continues heating the loop and thus keeps it warm longer, and the density is also slightly higher in the decay phase.

\subsection{Synthetic GOES and AIA Lightcurves}

 \begin{figure*}[t]
  \begin{center}
    \includegraphics[viewport =32 23 409 680, clip, angle=90,width=0.9\textwidth]{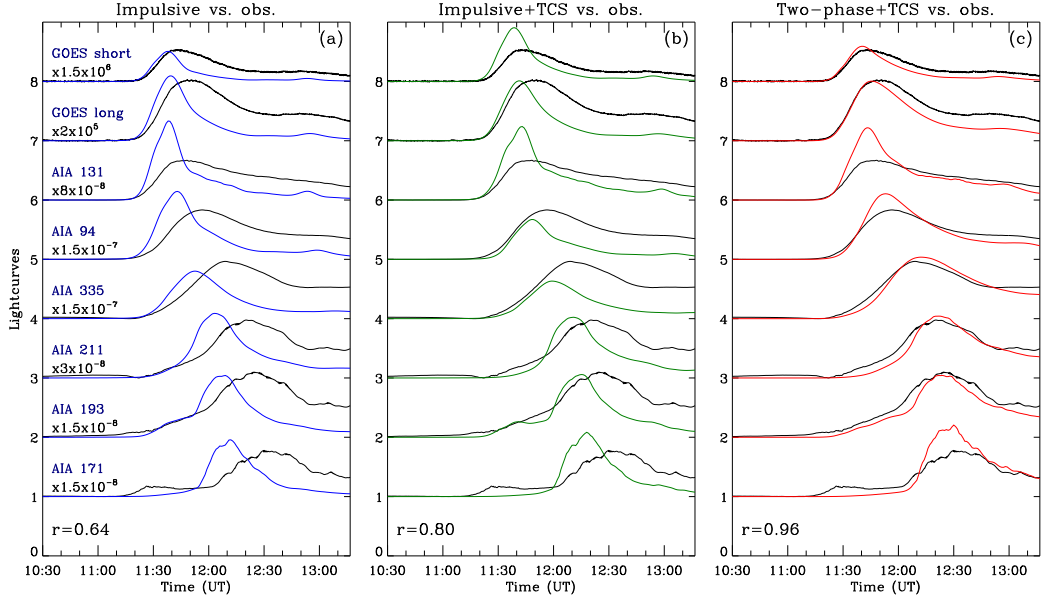}
    \caption{Comparison of the observed (black) and simulated (colored) lightcurves of the whole flaring region in two \GOES\ channels and six \SDO/AIA passbands. The background values are subtracted. Each flux is multiplied by the denoted factor and is offset by 1 from top to bottom. The three panels correspond to the results under three heating scenarios: impulsive heating (left), impulsive heating with TCS (middle), and two-phase heating with TCS (right). The average value of the correlation coefficients in each scenario is displayed at the lower-left corner in the corresponding panel. }
    \label{fig:fig3}
  \end{center}
\end{figure*}

With the evolution of each flaring loop modeled by EBTEL, the lightcurves of the whole flaring region are derived by convolving the Differential Emission Measure (DEM) calculated from multiple loops with the response functions of various channels from different instruments (\eg \citealp{qiu2012heating,liu2013determining,zeng2014flare,qiu2016long}). Figure~\ref{fig:fig3} gives the comparison of the synthetic lightcurves with the observations from \GOES\ soft X-ray and \SDO\ EUV channels, under those three heating scenarios, respectively.

With only impulsive heating and classical thermal conduction rate (see Figure~\ref{fig:fig3}(a)), the synthetic emission at high temperature $\ge$10~MK decays faster than observed, and the emission at 1~MK rises 20 minutes earlier than observed.  This indicates that the plasmas cool down faster in the simulation. In the impulsive heating with TCS scenario (Figure~\ref{fig:fig3}(b)), the cooling is delayed by $\sim$5 minutes, yet the difference between the model and observation is still remarkable. 
With only impulsive heating, the model cannot produce sufficient emissions at high temperatures after the peak of the flare, even though new heating events are still identified (Figures~\ref{fig:fig1}(c)\&(g)). 

With the inclusion of an extra slow tail in the heating function, \ie the two-phase heating with TCS displayed in Figure~\ref{fig:fig3}(c), the total flare emission at $\ge$~10~MK persists for a longer time with the lower temperature emission significantly delayed thereby agreeing with observations. This scenario produces sufficient emission in both the rise and decay phases of the flare, with the parameter set $\lambda_0 = 6.3{\times}10^5$ \Hunit, $\lambda_1 = 3.2{\times}10^5$ \Hunit, $\eta = 2.4{\times}10^6$ \etaunit. In Section 4.2, we discuss the rationale for the different choices of the scaling constant during the impulsive and gradual phases. 

To quantitatively evaluate the outputs of the three heating scenarios, the linear Pearson correlation coefficient in each channel (after comparable amplitudes obtained as shown in Figure~\ref{fig:fig3}) is calculated. The average values of those coefficients in each scenario, given in the lower-left corners (Figure~\ref{fig:fig3}), are 0.64, 0.80, and 0.96, respectively. This also suggests that the third scenario gives the best agreement to the observations.  Overall, the comparisons indicate that the flare might be involved with both TCS and two-phase heating.

\subsection{XRT and EVE Lightcurves}

 \begin{figure*}[t]
  \begin{center}
    \includegraphics[viewport =114 190 599 590, clip,angle=90, width=0.453\textwidth]{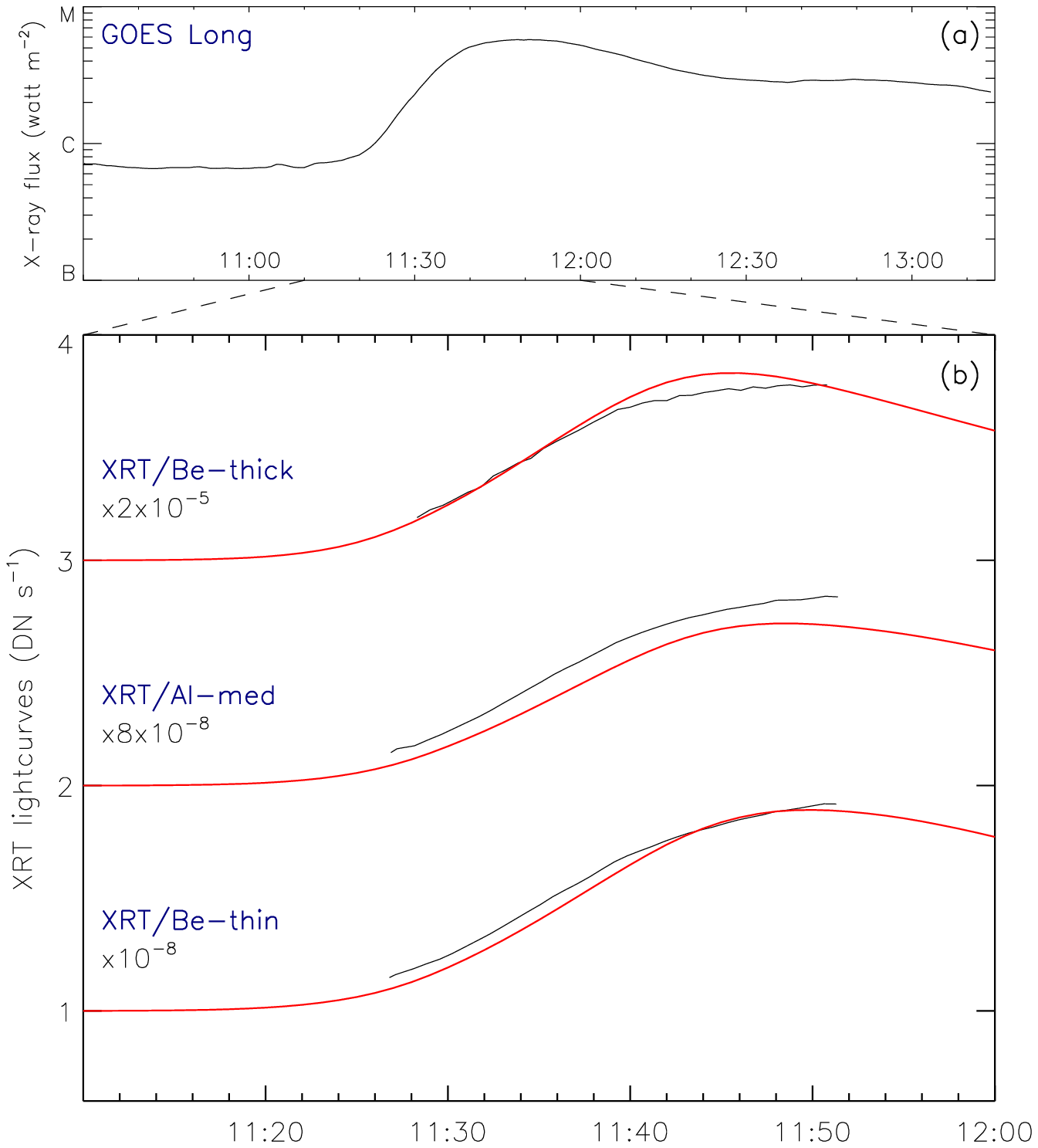}  
    \includegraphics[viewport =114 198 599 590, clip,angle=90, width=0.444\textwidth]{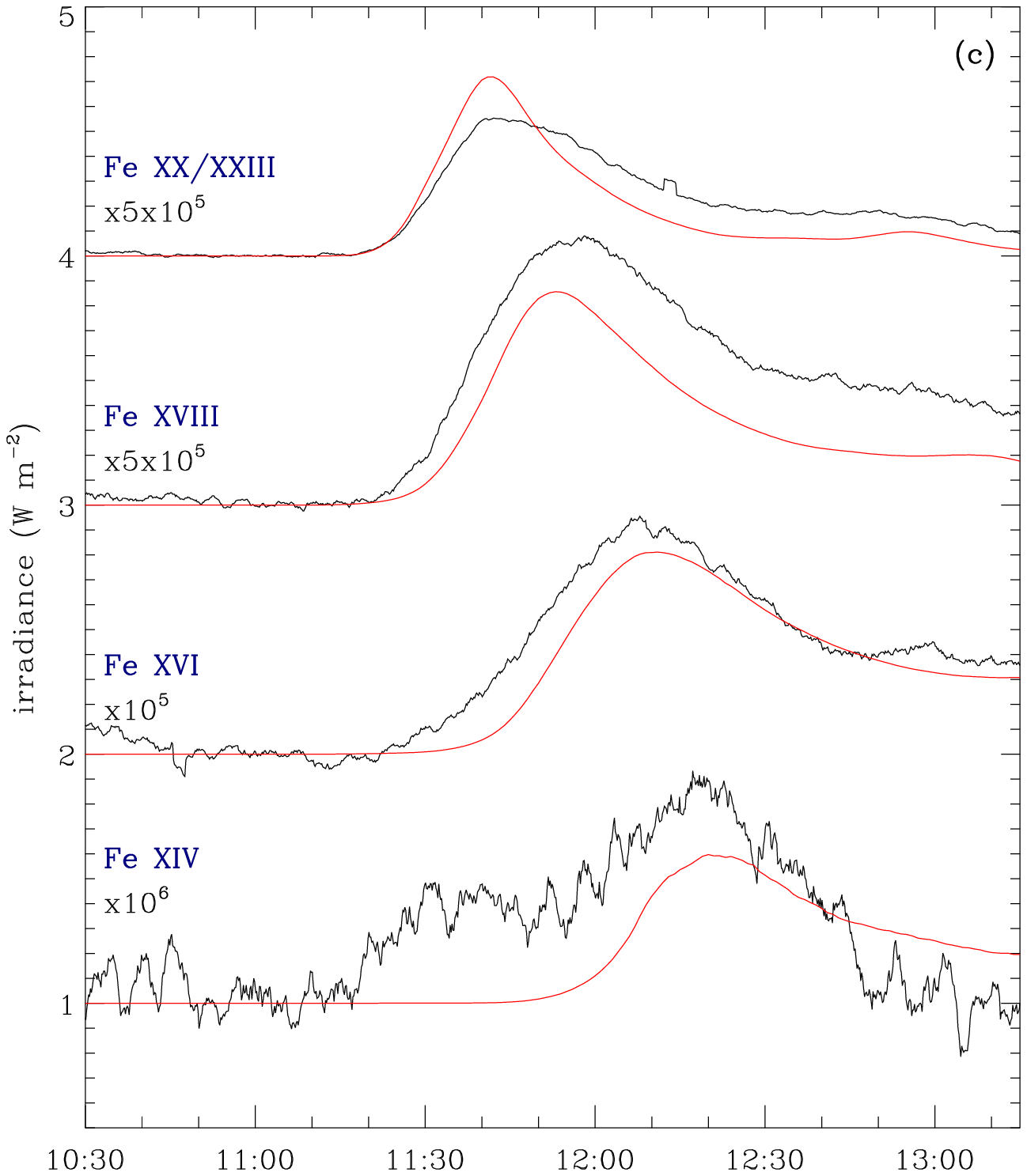}    
    \caption{\Hinode/XRT and \SDO/EVE lightcurves, compared with the synthetic ones given by EBTEL. The background values are subtracted. The observation is shown in black, while the simulation in red. Four typical channels in \SDO/EVE lightcurves are selected and shown in (c).}
    \label{fig:fig4}
  \end{center}
\end{figure*}

The two-phase heating model with TCS produces the synthetic X-ray and EUV light curves in reasonable agreement with the GOES and AIA observations; therefore, we use this model and make further comparisons of the synthetic lightcurves to the observed X-ray flux from \Hinode/XRT and EUV lines from \SDO/EVE, as displayed in Figure \ref{fig:fig4}. 

Figures~\ref{fig:fig4}(a){\&}(b) give the \Hinode/XRT coverage of this flare between 11:27--11:51~UT, roughly corresponding to the early phase until the flaring peak. Three XRT channels are listed in Figure \ref{fig:fig4}(b), including Be-thick, Al-med, and Be-thin. As there is no data covering this region before the flare, the background level in each channel is estimated with the average value of pixels outside the flaring region. Then the total background contribution is subtracted from the original lightcurves. The results in Figure~\ref{fig:fig4}(b) suggest the two-phase heating gives good agreement with the observations in those three channels.

Figure \ref{fig:fig4}(c) shows the observed and synthetic EVE curves during the flare. They also display good agreement in the listed typical emission lines including \ion{Fe}{20}/{\small \Rmnum{23}} ($\log$T $\sim$6.97), \ion{Fe}{18} ($\log$T $\sim$6.81), \ion{Fe}{16} ($\log$T $\sim$6.43), and \ion{Fe}{14} ($\log$T $\sim$6.27), with both comparable peaking values and decay time. Though the observed cooler  \ion{Fe}{14} line has complicated profiles possibly due to other contributions such as the emissions from the transition region.   

\subsection{DEMs}

 \begin{figure*}[t]
  \begin{center}
    \includegraphics[viewport =51 145 759 477, clip, width=0.9\textwidth]{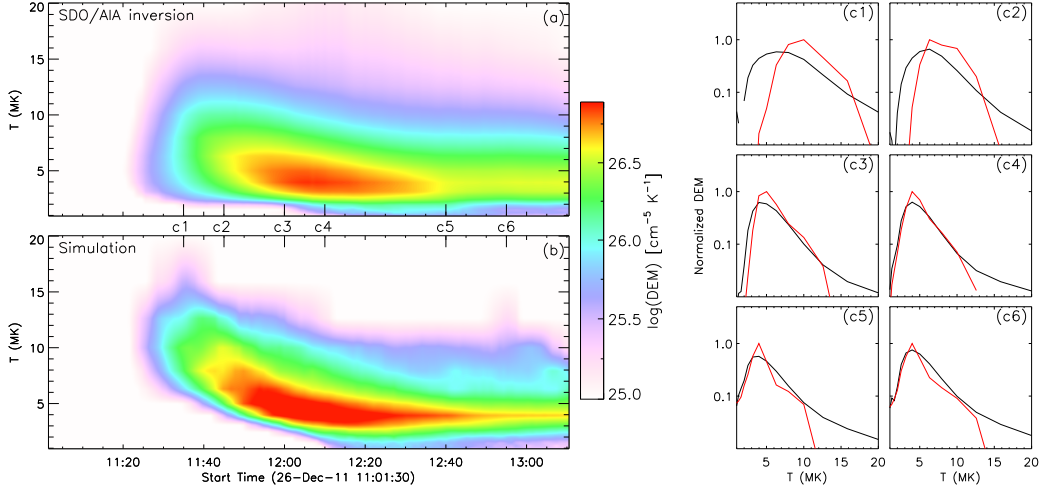}   
    \caption{Comparison of the \SDO/AIA inverted and synthetic DEM distributions applying the two-phase heating with TCS method. The same color scale is used. The background level, chosen from 11:10--11:15 UT, is subtracted. The DEMs at six times c1--c6 are indicated in the right panels accordingly. The black and red curves correspond to the AIA inverted and synthetic DEMs, respectively.}
    \label{fig:fig5}
  \end{center}
\end{figure*}

The distributions of the DEMs covering the whole flaring region are inverted from the \SDO/AIA observation and also are synthesized from the EBTEL simulation, as shown in Figures~\ref{fig:fig5}(a){\&}(b), respectively. The sparsity-based inversion method for the DEMs by \citet{cheung2015thermal} is used for this inversion. Figure \ref{fig:fig5}(b) gives the synthetic DEMs under the scenario of two-phase heating with TCS. For both DEM maps, the averaged values from 11:00--11:15~UT are chosen as the background levels and thus get subtracted from the original DEM values. Both maps display a downward trend before $\sim$12:30~UT and stays roughly flat thereafter, and both give higher peaking DEM values during 11:45--12:30~UT. A clear difference is that the DEMs inverted from the observation have broader distributions than the simulation, and the former has more contribution from plasma hotter than $\sim$12 MK. 

The DEMs at six times, as indicated between Figures \ref{fig:fig5}(a){\&}(b) by c1--c6, are shown in Figures \ref{fig:fig5}(c1--c6) accordingly. Before $\sim$12:00~UT including c1 and c2, the peaks of the simulated DEMs are higher and shift to hotter temperatures by a few MK than the observationally inverted ones. After that, the peaks of those DEMs are comparable in the magnitude and also the associated temperatures. Besides, larger inverted DEMs at very hot temperatures ($>$12 MK) are also noticeable in those profiles.

These comparisons indicate that the EBTEL well reveals the general evolution of the DEM during this flare, especially in its decay phase. Though little emission from a temperature larger than 12 MK are present in the result of EBTEL, which might be due to the 0D nature of EBTEL based on the average values of the loops.   

\section{Energetics of the Flare}
\subsection{Energy Partition}

     \begin{figure*}[t]
  \begin{center}
    \includegraphics[viewport =63 173 919 477, clip,width=0.9\textwidth]	{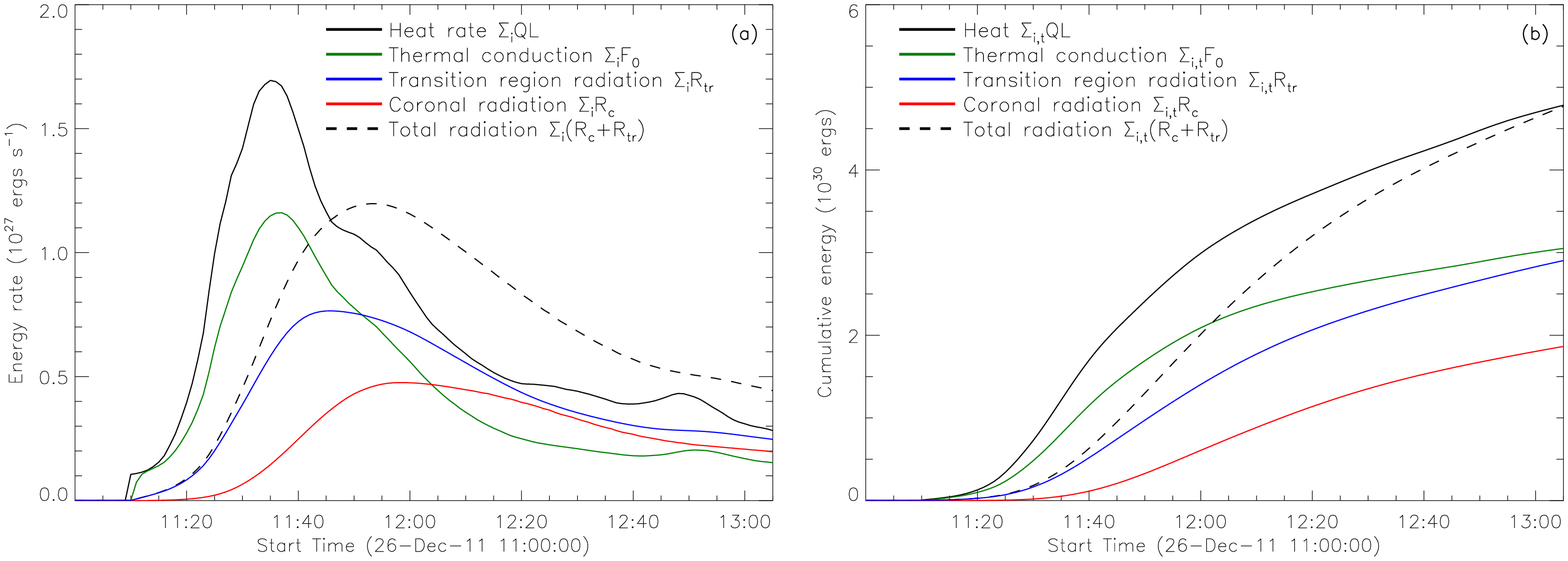}
    \caption{Evolution of the estimated energy rates (a) and cumulative energies (b) from the EBTEL simulation using the two-phase heating with TCS mechanism.}
    \label{fig:fig6}
  \end{center}
\end{figure*}

The evolutions of the total heating rate and the cumulative heating energy are estimated and displayed in  Figure~\ref{fig:fig6}. The peak of heating rate is $1.7{\times}10^{27}$ ergs~s$^{-1}$ at 11:35 UT. With the increased temperatures of the flaring loops due to the impulsive heating, the thermal conduction increased accordingly and peaks at 11:37 UT with ${\sim}1.2{\times}10^{27}$ ergs~s$^{-1}$.  As the temperature tends to increase earlier than the density, as evident from Figures \ref{fig:fig2}(c{\&}d), the peak of the transition region radiation ($\mathrm{R_{tr} \varpropto p \varpropto nT}$) appears earlier than coronal radiation ($\mathrm{R_c \varpropto n^2}$). The peaking values of $\mathrm{R_{tr}}$ is $7.7{\times}10^{26}$ ergs~s$^{-1}$ at 11:46 UT,  around 1.6 times the peaking $\mathrm{R_c}$ of $4.8{\times}10^{26}$ ergs~s$^{-1}$ at 11:58 UT. 
  
To study the energy partitions during the flare, the cumulative energies are tracked and shown in Figure \ref{fig:fig6}(b).  By 13:10 UT, the total heat input is around $4.9{\times}10^{30}$ ergs, roughly balanced by the total radiation energy which is composed of $\mathrm{R_{tr\_tot}}$ of $3.0{\times}10^{30}$ ergs and $\mathrm{R_{c\_tot}}$ of $1.9{\times}10^{30}$ ergs. The coronal radiation can also be estimated from the GOES soft X-ray data (\citealp{Cox1969,Emslie2005}). It gives a value of $2.2{\times}10^{30}$ ergs, roughly agrees with $\mathrm{R_{c\_tot}}$ from our simulation. The total thermal conduction loss is roughly at $3.1{\times}10^{30}$ ergs, which is radiated through the transition region.  

\begin{figure*}[t]
  \begin{center}
    \includegraphics[viewport =248 87 418 670, clip, angle=90,width=0.98\textwidth]	{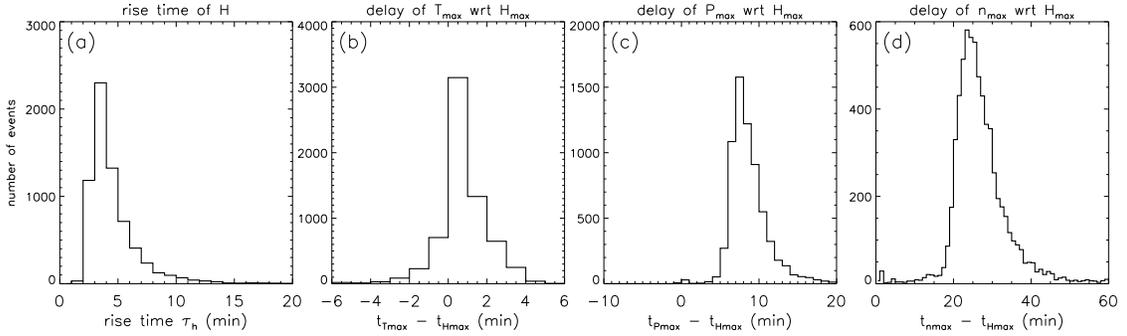}
  \caption{Distributions of the rise times of the heating rates (a), and the delays of the peak values of the temperature ($\mathrm{T_{max}}$) in (b), pressure ($\mathrm{P_{max}}$) in (c), and density ($\mathrm{n_{max}}$) in (d) with regard to the peak heating rate $\mathrm{H_{max}}$ in each loop. }
    \label{fig:fig7}
  \end{center}
\end{figure*}

\begin{figure*}[t]
  \begin{center}
    \includegraphics[viewport =145 84 409 675, clip, angle=90,width=0.75\textwidth]	{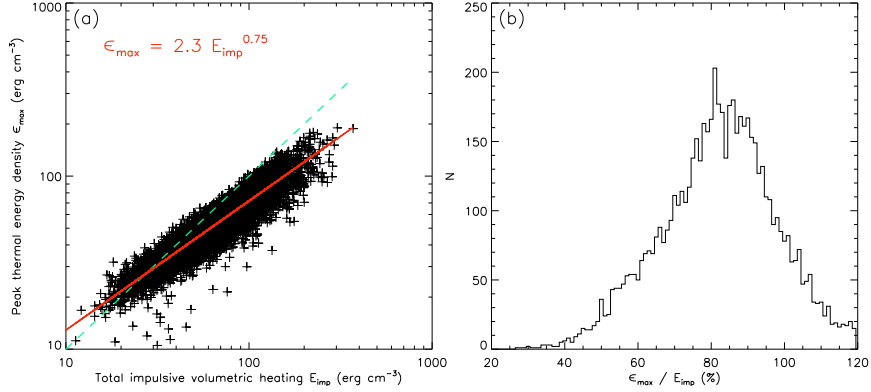}
  \caption{Relationship of the peak thermal energy density $\mathrm{\epsilon_{max}}$ in each loop to its total impulsive volumetric heating $\mathrm{E_{imp}}$. (a) Their power law relation is indicated by the solid line in red, with a fitting shown in the top-left corner. The dashed line indicates positions along y = x. (b) The histogram of  $\mathrm{\epsilon_{max}/E_{imp}}$.   }
    \label{fig:fig8}
  \end{center}
\end{figure*}

\subsection{Energetics in the Two Phases}
Assuming that a half flare loop is anchored at each UV brightened AIA pixel, we have identified and modeled over 6,700 half loops, each with a different heating rate and length as constrained by observations.
In this study, each flare loop is heated ``impulsively'' and then gradually, as demonstrated by the two-phase UV light curve at the foot-point. We explore the different roles of the heating in the two phases. 

Figure~\ref{fig:fig7}(a) shows the distribution of the rise times of the UV light curves. The rise times primarily range between 2 to 6 minutes. The timescale of thermal conduction 
$\tau_{cond}$ using the Spitzer thermal conductivity at temperature 1--10 MK and density 10$^{9-10}$ cm$^{-3}$ is no longer than 1 minute, and 
the reaction of the transition region to energy deposition is of order a few seconds. Therefore, the observed rise time of the UV light curves is substantially longer than the timescale of thermal conduction, indicating that the observed rise time is characteristic of the heating timescale. 
We also note that for the coronal plasma at temperature 1--10 MK and the length of the coronal loop at 30~Mm, the characteristic acoustic time is 1--3 min, which is a fraction of the heating timescale. If the AIA instrument, at the resolution of 0.6\arcsec, nearly resolves individual flare loops, then flare loops would mostly evolve in quasi-equilibrium even during the ``impulsive" phase. If an AIA-identified flare loop consists of sub-structures like threads, the heating time of each thread could be shorter (\citealp{graham2015temporal}).

The next three panels in Figure~\ref{fig:fig7} show the distributions of the time lags of the peak temperature, pressure, and density of a loop relative to its time of the peak heating rate $\mathrm{\tau_1 = t_{Tmax} - t_{Hmax}}$, $\mathrm{\tau_2 = t_{Pmax} - t_{Hmax}}$, and $\mathrm{\tau_3 = t_{nmax} - t_{Hmax}}$, respectively. It is seen that the temperature of the corona peaks shortly after the peak heating rate, whereas the pressure peaks a few minutes later, when the impulsive heating has nearly finished. These results indicate that the impulsive heating raises the thermal energy of the coronal loop, so that the thermal energy density $\epsilon$ is roughly proportional to the time integral of the volumetric heating rate $Q$, 
$\mathrm{\epsilon_{max} = (3/2)P_{max} \sim \int Q_{imp} dt}$. Figure~\ref{fig:fig8} further corroborates this point. Figure~\ref{fig:fig8}(a) gives the scatter plot of  $\mathrm{\epsilon_{max}}$ versus $\mathrm{E_{imp} = \int Q_{imp} dt}$, showing that the two are scaled, though not exactly by a linear relation, due to a certain amount of radiative loss. Figure~\ref{fig:fig8}(b) shows the ratios of $\mathrm{\epsilon_{max}}$ to $\mathrm{E_{imp}}$. The mean ratio is about 80\%, suggesting that most of the impulsive heating energy is used to raise the thermal energy of the corona loop, and the rest 20\% is lost by radiation.

Figure~\ref{fig:fig7}(d) displays the lag of the peak density (as well as the peak coronal radiative loss) relative to the time of the peak heating rate, which is about 10 minutes later than $\mathrm{P_{max}}$. Therefore, during the phase of impulsive heating, the coronal radiative loss can be ignored. With these results, it is seen that, during the impulsive heating, the energy equation is reduced to $\mathrm{Q \approx |dP/dt| + |R_{tr}/L|}$.

That the coronal pressure $\mathrm{P}$, or thermal energy density  $\mathrm{\epsilon = (3/2) P}$, reaches the maximum at the end of the impulsive heating, suggests that in the gradual phase, the heating energy is at most used to balance the radiative loss and does not continue to increase the thermal energy of the flare loop. In this phase, the coronal radiation becomes important, whereas the coronal pressure varies slowly. Therefore, in this phase, the energy equation is approximately $\mathrm{Q \approx (|R_c| + |R_{tr}|)/L}$. In our empirical model, we infer heating rates of flare loops from (transition region) UV radiation by a scaling factor $\lambda_0$ during the impulsive heating phase and $\lambda_1$ during the slow-heating phase. Different governing physics during these two phases specifies different relations between the heating rate and transition region radiation, which may explain why $\lambda_0$ is different from $\lambda_1$.

\begin{figure*}[t]
  \begin{center}
    \includegraphics[viewport = 74 30 415 610, clip, angle=90,width=0.9\textwidth]	{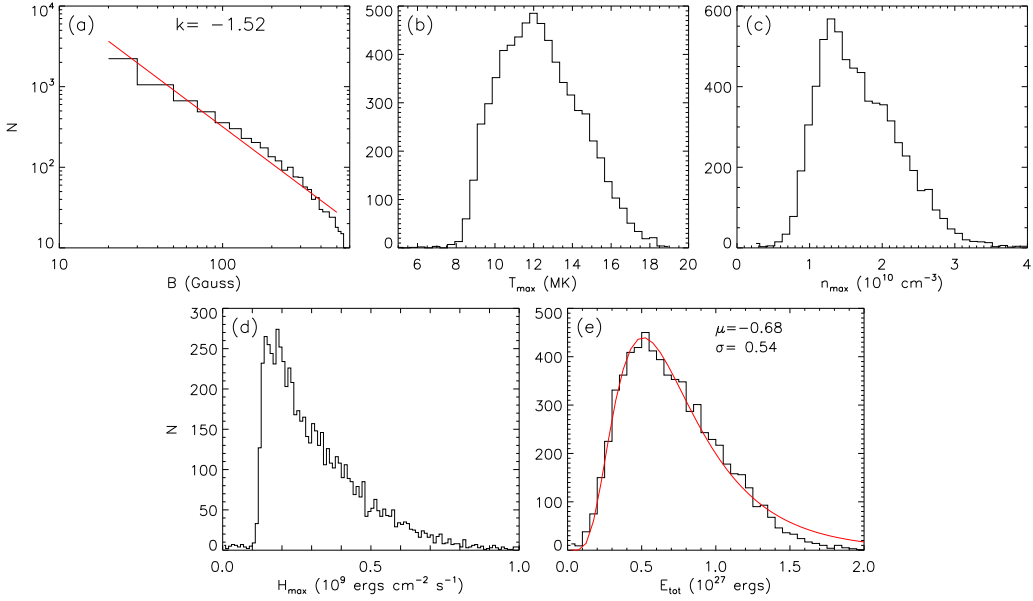}
  \caption{Histograms of the physical parameters in the flaring loops. (a) The magnetic field strength at the loop footpoints. A power-law fitting is indicated by a red line, giving a power law index of -1.52. (b) Peak temperature $\mathrm{T_{max}}$. (c) Peak density $\mathrm{n_{max}}$. (d) Peak heating rate $\mathrm{H_{max}}$. (e) Total energy release $\mathrm{E_{tot}}$. A log-normal fitting is denoted by a red curve, with its center $\mu$ and width $\sigma$ given at the top. } 
    \label{fig:fig9}
  \end{center}
\end{figure*}

\subsection{Properties of the Flare Loops}

We also examine the distribution of physical parameters of these 6,700 half loops. Figure~\ref{fig:fig9} shows the histograms of the magnetic field strength, peak temperature, peak density, peak heating flux, and the total heating energy of these loops. The magnetic field strengths at the loop footpoints have a power-law distribution, with an index of -1.52. The peak temperature ranges from $\sim$8--18 MK. In this flare, the peak heating flux ranges from 10$^{8-9}$ ergs~cm$^{-2}$~s$^{-1}$. Heating flux of this order usually does not generate a strong chromosphere evaporation (\citealp{fisher1985flare,reep2015}); as a result, the peak density of this flare is of order 1--3$\times$10$^{10}$~cm$^{-3}$. 

The total heating energy in the flare loop ranges between 10$^{26-27}$
ergs, or each flare loop is equivalent to a micro-flare (\citealp{hannah2011microflares}). 
In this flare, the distribution of the total energy released in each flare loop can be fitted to a log-normal distribution (Figure \ref{fig:fig9}(e)). The center $\mu$ and width $\sigma$ of this fitting are -0.68 and 0.57, respectively. Here the total energies follow a log-normal distribution, possibly related to the similar distribution of the magnetic flux concentrations (\citealp{abramenko2005distribution}).

\begin{figure*}[t]
  \begin{center}
    \includegraphics[viewport = 110 60 580 648, clip, angle=90,width=0.7\textwidth]	{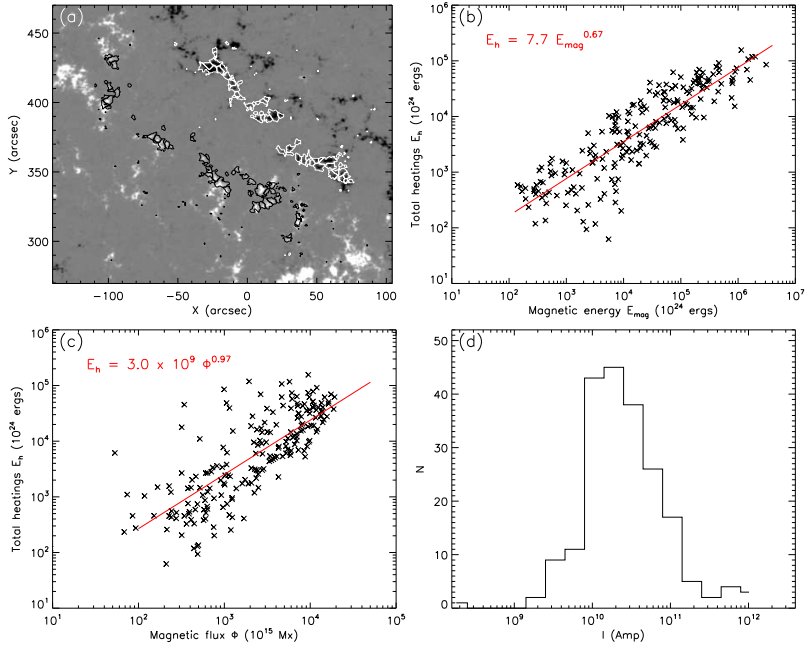}
  \caption{Scalings of the heatings with the magnetic energies and fluxes in the magnetic concentrations. (a) The magnetic concentrations in the flaring regions. Their boundaries are outlined by black/white curves around positive/negative magnetic fields. (b) The total heatings $\mathrm{E_{h}}$ versus the total magnetic energies $\mathrm{E_{mag}}$ in the flux concentrations. A power-law is indicated by the solid line, with the relationship given at the top left. (c) $\mathrm{E_{h}}$ versus the magnetic flux ($\mathrm{\Phi}$) in the concentrations. The denotation is similar to (b). (d) The histogram of the estimated strengths of the electric currents in the current sheets that are associated with the concentrations during the flare. }  
    \label{fig:fig10}
  \end{center}
\end{figure*}

\subsection{Properties of the Flux Concentrations} 
We look into the relationships of the total heatings with magnetic energies and fluxes in the flaring flux concentrations, as seen in Figure~\ref{fig:fig10}. The flaring regions are partitioned into  flux concentrations using the method presented by \citet{abramenko2005distribution}. Here the flaring locations with magnetic field strength larger than a threshold of 25 Gauss are considered.  This accounts for 71\% of the 6,700 flaring pixels. There are 206 flux concentrations identified and outlined in Figure~\ref{fig:fig10}(a).

Figure~\ref{fig:fig10}(b) shows the scaling of the total heating $\mathrm{E_h}$ with the magnetic energy $\mathrm{E_{mag}}$ in the magnetic flux concentrations. $\mathrm{E_{mag}}$ is estimated to be $\mathrm{{\sum}_i (1/8{\pi}){B_i}^2{L_i}S}$, where $\mathrm{B_i}$ and $\mathrm{L_i}$ are the footpoint field strength and the length of loop i in a chosen concentration, respectively, and S is the area of one pixel. In this event, $\mathrm{E_h}$ and $\mathrm{E_{mag}}$ can be scaled with a power law, \ie $\mathrm{E_h = 7.7 {E_{mag}}^{0.67}}$. Overall, the amount of the heating energy is 12\% of the magnetic energy calculated in flaring pixels. 

Similar procedure is applied to check the relationship of $\mathrm{E_h}$ and the magnetic flux $\mathrm{\Phi}$, giving an equation of $\mathrm{E_h = 3.0{\times}10^9 {\Phi}^{0.97}}$, with $\mathrm{E_h}$ in unit of ergs and $\mathrm{\Phi}$ in Mx,  as shown in Figure~\ref{fig:fig10}(c). The nearly linear relationship between the flare heating energy and magnetic flux suggests that the two physical quantities are scaled by the mean electric current in the current sheets $\mathrm{\langle I \rangle \sim 3\times 10^{10}}$ Amp.
The distribution of this current $\mathrm{I =E_h/\Phi }$ in the current sheet(s) associated with each flux concentration is shown in  Figure~\ref{fig:fig10}(d). It is in order of $10^{10}$ Amps, consistent with previous studies (\eg \citealp{longcope2007modeling,qiu2009observational,longcope2010quantitative}).

\section{Discussion and Conclusions}

We modeled a typical two-ribbon flare of C5.7 class on 2011 December 26 observed by \SDO, \Hinode\ and \GOES\ to determine the heating rates in $\sim$6,700 half flaring loops. Three heating scenarios are tested with the 0D EBTEL model, including impulsive heating, impulsive heating with TCS, and two-phase heating with TCS, among which the latter gives the best agreement with the observed X-ray and EUV light-curves.

The peak temperatures and densities of the flaring loops are around 12 MK and $\mathrm{1.5\times 10^{10} ~cm^{-3}}$ (Figures~\ref{fig:fig9}), respectively, which imply that the thermal flux can be locally limited (\eg \citealp{battaglia2009observations}). In this study, the minimum values of the TCS reduction factors for each half loop are among 0.07--0.13.  The TCS results in a higher coronal temperature (Figure~\ref{fig:fig2}(c)), so that the simulation better agrees with observations at hot channels (Figures~\ref{fig:fig3}{\&}\ref{fig:fig4}). However, our study suggests that the impulsive heating with TCS cannot reproduce the observed slow cooling process in this flare (Figures~\ref{fig:fig3}), and an additional persistent low-rate heating is necessary to agree with observations. This result is consistent with the recent study by \citet{bian2018heating}, which suggests that both the extended duration of magnetic energy release and the suppression of heat conduction are needed to explain the inferred physical properties from flare observations.   

Under the two-phase heating scenario, the total input energy is composed of impulsive and gradual heating with amounts of $2.8{\times}10^{30}$ and $2.1{\times}10^{30}$ ergs, respectively, \ie the impulsive and slow heating components account for  60\% and 40\% respectively of the total heating during this flare. The timescale of the impulsive heating, as inferred from the observed UV light curves, ranges between 2--6 minutes, and that of the ensuing slow heating is typically over 20 minutes, considering the decay time scale of the lightcurves (\eg \citealp{qiu2010reconnection, cheng2011hard, liu2013determining}). The peak heating flux in the impulsive phase reaches a few times 10$^8$ ergs~cm$^{-2}$~s$^{-1}$, and during this phase, the heating energy is mostly used to raise the thermal energy of the coronal loop. The slow-heating at a lower rate, of a few times 10$^{7}$~ergs~cm$^{-2}$~s$^{-1}$, does not increase the thermal energy of the loop, and nearly balances the radiative losses in the corona as well as the transition region, allowing the loop to cool more gradually than otherwise. The observed slow decay of the UV light curve at the foot-point of a flare loop is a reflection of the slowly evolving corona, which keeps heating the transition region by thermal conduction. Previously, \citet{liu2013determining} modeled the flare loop evolution using only an impulsive heating, and calculated the foot-point UV radiation caused by thermal conduction of the corona without additional heating during the decay phase. They found that the synthetic flux of \ion{C}{4}, which is dominating in the 1600 \AA ~emission during flares, can roughly account for around half of the observed values in this channel during the long decay. In this paper, we illustrate the need for additional heating in the decay phase of a flare loop, which increases the thermal conduction and therefore the UV emission in the decay phase. Several possible explanations for the slow heating process were proposed (see \citet{qiu2016long} and references therein). One observational constraint to those theories could be the duration of this process, which is roughly at 20--40 minutes (Figures \ref{fig:fig2}(a){\&}(b)).

Based on the evolution of plasmas in the flaring loops, the energy partitions are estimated. Our calculations indicate that the total heating and radiation energies for this C5.7 flare are roughly equivalent, both at a level of $4.9{\times}10^{30}$ ergs. The total kinetic energy of the associated CME is estimated to be $1.2{\times}10^{31}$ ergs\footnote{\url{https://cdaw.gsfc.nasa.gov/CME_list/UNIVERSAL/2011_12/univ2011_12.html}}. So, for this eruption, the energies distributed in the flare and CME is comparable, agreeing with the previous conclusions (\eg \citealp{Emslie2005}). The cumulative thermal conduction of $3.1{\times}10^{30}$ ergs, is roughly balanced by the radiation from transition region, indicating that most of the thermal energies conducted from the corona are finally dissipated in the lower atmosphere. 

EBTEL is a powerful tool to investigate the evolution of the coronal loops/threads, yet its limitations and also the assumptions in this study need to be considered. Many of those have been discussed by the recent work of \citet{qiu2016long}. Some dynamic processes in each flaring loop, such as the observed shrinkage affecting its length (\eg \citealp{savage2011quantitative, zhu2016observation}), usually last for a few minutes or less, which is small compared to the whole flaring timescale. Thus EBTEL is expected to provide a good approximation at least in the long gradual phase of the flare. Other effects, \eg how the spatial and temporal changes of the cross section (\citealp{klimchuk2001cross,mikic2013importance}) and the plasma composition (\citealp{phillips2004solar,Barnes2016}) affect the hydrodynamic evolution of a loop/thread should be evaluated in the future study. 

For the future work, we will look further into the role of the slow heating in solar flares to answer some related questions such as, whether it is ubiquitous in the flares, how much it varies with different magnitudes of flares, and what are the mechanisms for impulsive and slow heatings in a flare. We will also investigate the heating process for more flares with complex configurations, and see how it may vary with the evolving magnetic structures.

\acknowledgments The authors thank the referee for several constructive comments. This work is supported by the NASA grant NNX14AC06G, the NSF SHINE collaborative grant AGS-1460059, and the ISSI/ISSI-BJ team ``Diagnosing Heating Mechanisms in Solar Flares". We thank Sarah Pearce for the preliminary study.
\facilities{\SDO, \Hinode/XRT, \GOES}

\bibliographystyle{yahapj} 
\bibliography{twophase}		

\end{document}